\documentclass[lettersize,journal]{IEEEtran}
\IEEEoverridecommandlockouts
\usepackage{cite}
\usepackage{amsmath,amssymb,amsfonts}
\usepackage{algorithmic}
\usepackage{graphicx}
\usepackage{textcomp}
\usepackage{xcolor}
\usepackage{ulem}

\def\BibTeX{{\rm B\kern-.05em{\sc i\kern-.025em b}\kern-.08em
    T\kern-.1667em\lower.7ex\hbox{E}\kern-.125emX}}
\begin{document}

\title{
Energy-Aware 6G Network Design: A Survey

}

\author{\IEEEauthorblockN{
 Rashmi Kamran\IEEEauthorrefmark{1},
 Mahesh Ganesh Bhat\IEEEauthorrefmark{2}, 
 Pranav Jha\IEEEauthorrefmark{1},
 Shana Moothedath\IEEEauthorrefmark{2},
 \\Manjesh Hanawal\IEEEauthorrefmark{3},
 Prasanna Chaporkar\IEEEauthorrefmark{1} 
\\\IEEEauthorblockA{Department of Electrical Engineering,
Indian Institute of Technology Bombay, India\IEEEauthorrefmark{1}\\
\IEEEauthorblockA{Department of Electrical and Computer Engineering, Iowa State University, Ames, IA, US\IEEEauthorrefmark{2}\\
\IEEEauthorblockA{Department of Industrial Engineering and Operations Research, Indian Institute of Technology Bombay, India\IEEEauthorrefmark{3}\\
Email: 
rashmi.kamran@iitb.ac.in\IEEEauthorrefmark{1}, mgbhat@iastate.edu\IEEEauthorrefmark{2},
pranavjha@ee.iitb.ac.in\IEEEauthorrefmark{1},
mshana@iastate.edu\IEEEauthorrefmark{2},
\\mhanawal@iitb.ac.in\IEEEauthorrefmark{3}, chaporkar@ee.iitb.ac.in\IEEEauthorrefmark{1} }}}}}
\maketitle

\begin{abstract}
 6th Generation (6G) mobile networks are envisioned to support several new capabilities and data-centric applications for unprecedented number of users, potentially raising significant energy efficiency and sustainability concerns. This brings focus on sustainability as one of the key objectives in the their design. To move towards sustainable solution, research and standardization community is focusing on several key issues like energy information monitoring and exposure, use of renewable energy, and use of Artificial Intelligence/Machine Learning (AI/ML) for improving the energy efficiency in 6G networks. The goal is to build energy-aware solutions that takes into account the energy information resulting in energy efficient networks. Design of energy-aware 6G networks brings in new challenges like increased overheads in gathering and exposing of energy related information, and the associated user consent management. The aim of this paper is to provide a comprehensive survey of methods used for design of energy efficient 6G networks, like energy harvesting, energy models and parameters, classification of energy-aware services, and AI/ML-based solutions. The survey also includes few use cases that demonstrate the benefits of incorporating energy awareness into network decisions. Several ongoing standardization efforts in 3GPP, ITU, and IEEE are included to provide insights into the ongoing work and highlight the opportunities for new contributions. We conclude this survey with open research problems and challenges that can be explored to make energy-aware design feasible and ensure optimality regarding performance and energy goals for 6G networks.
\end{abstract}

\begin{IEEEkeywords}
Energy awareness, Energy information exposure, Energy efficient networks, Sustainable network design, Policy update 
\end{IEEEkeywords}

\section{Introduction}
\par 
The emergence of mobile network technologies not only cater to the connectivity needs of mobile users, but their applications cater to a vast number of industry and enterprise domains such as mining, oil and gas sectors, utilities, healthcare, education, manufacturing, agriculture, and many more \cite{itu-M2527}.
In addition, there are many new upcoming use cases and applications for future mobile network technologies that require an extensive set of capabilities, such as holographic telepresence, unmanned mobility, immersive tourism, and precision planting \cite{ambar}, \cite{wei}. Future mobile networks need enhancements in their capabilities (such as data rate, spectrum efficiency, connection density, latency, and reliability) and additional capabilities (such as ubiquitous coverage, positioning, sensing-related, and Artificial Intelligence (AI)\,/\,Machine Learning (ML)-related capabilities, sustainability, interoperability) to fulfill the technical and operational aspects of a large pool of applications for a massive number of users\cite{itu-M2160}, \cite{wang++2023}. 
Worldwide mobile subscribers are expected to be 9.47\,billion, and mobile data traffic per active smartphone is expected to be 40\,Gb per month by 2030, which was 19\,Gb per month in 2024. Rising demand for a range of communications services catering to billions of users and devices (for example IoT devices) along with data-centric AI/ML integration results in energy consumption versus performance trade-offs in mobile networks  \cite{sabella}, \cite{zhang},  \cite{chou2024}. 
\par In this context, a framework and overall objectives for developing International Mobile Telecommunications (IMT) 2030 are provided in an International Telecommunication Union (ITU) recommendation \cite{itu-M2160}. The recommendation ITU-R M.2160 contains the guidelines for defining the Sixth Generation (6G) mobile network design principles \cite{ITU-press}. Sustainability is one of the highlights in the clauses of ``\textit{Motivation and societal considerations}" and ``\textit{User and application trends}" in this recommendation. In addition, environmental impact, energy efficiency, and automated network programmability are also highlighted as value indicators for 6G by Next Generation Mobile Networks (NGMN) Alliance \cite{ngmn}. Hence, an energy-aware design considering ubiquitous connectivity and ubiquitous intelligence deployments is a key requirement for future mobile networks towards achieving sustainability goals\cite{zhao++}, \cite{ziegler2020}, \cite{bolla2023}, \cite{shehab2022}, \cite{govindasamy2023}.
\par 
For achieving sustainability goals in future mobile networks, energy-aware   service delivery in mobile networks is the need of the hour. 
The increased usage of renewable energy in mobile networks as well as use of energy as per need without any over provisioning can make the networks more sustainable. 
Additionally, users having a choice to lower their
energy consumption, even with compromised service performance, or opting for renewable energy, can reduce their carbon footprint.
Hence, a coordinated solution between the users and the network is required to reduce energy consumption in intelligent mobile networks and to prioritize the usage of renewable energy. To explore this perspective in detail, our survey paper focuses on energy-aware   design concepts and their implications for 6G networks, including use cases, benefits, challenges, and standardization initiatives in this direction.
\begin{table*}[h]
  \centering
  \caption{Existing Survey}
  \label{survey}
  \begin{tabular}{|p{1.2cm}|p{14.7cm}|}
  \hline
    Literature 
    & Domains covered \\
     \hline \cite{luo2025-survey} 
& Symbiotic blockchain 6G network, sharding scheme, energy analysis \\
     \hline \cite{li2025-survey} 
     & AI/ML model deployment for energy efficiency, Multi-tiered cloud-enabled endogenous intelligent architecture for 6G wireless networks \\
     \hline \cite{ahmadi2025-survey} 
& Sustainability requirements for 6G, Open RAN as a key architectural solution \\
 \hline \cite{dbouk2025-survey} 
 & AI energy-saving-based techniques for 5G base stations, Thermal Management of Heat Transfer in 5G Networks \\
 \hline \cite{you2025-survey} 
 & Impact of Ubiquitous AI on sustainability goals for 6G, Lightweight AI techniques \\
  \hline \cite{singh2025-survey} 
  & Role of SDN, NFV, and C-RAN in creating a sustainable and green cellular network, framework for energy efficient cellular networks \\
  \hline \cite{lorincz2024-survey} 
& 5G network slicing and the energy demand of different network
slicing use cases, key performance indicators for the assessment of 5G network slices,
different implementations and approaches to network slicing for energy saving in the network\\
 \hline \cite{pivoto} 
& 6G use case and enabling technologies, methods and variables for relevance analysis\\

\hline \cite{larsen2023-survey} 
& Modeling of the RAN energy consumption, Energy efficient RAN architectures: virtual RAN, cloud RAN, and open RAN, Network sharing from energy perspective \\
\hline \cite{ali2018-survey} 
& Energy efficiency in HetNets and HCRANs, Resource management and network optimization, RRH clustering, Energy efficient cloud computing\\
    \hline \cite{buzzi2016-survey} 
    & Resource allocation, network planning and deployment, energy harvesting and transfer, and hardware solutions \\
    \hline \cite{muhammad2015-survey}
    & Modeling of energy efficiency in wireless network, traffic modeling and performance metrics, Green solutions: BS on off switching, MT radio interface on-off switching, scheduling for multi-homing access and with multiple energy sources\\
    \hline \cite{feng2013-survey} 
    & EE metric consumption models, radio resource optimization, energy efficient network deployment strategies, MIMO and cross layer design for energy efficiency \\
    \hline \cite{wang2012-survey}
    & Over provisioning, Energy performance trade-off, ON/OFF resource allocation and virtualization of data center resources, cell zooming, power saving of power amplifiers, green metrics: equipment level metrics and facility level metrics, prediction based green techniques\\
\hline \cite{hasan2011-survey} 
& Architectures for energy savings in base stations, Network planning for heterogeneous network deployment, Cognitive radio and cooperative relaying \\
    \hline
  \end{tabular}
    \label{tab:1}
\end{table*}

\subsection{How this Survey differs from the Existing Surveys}


Several survey papers have considered energy-related aspects of mobile networks.
~\cite{buzzi2016-survey} highlights energy-efficient techniques for Fifth Generation (5G) mobile networks with a scope limited to resource allocation, network planning and deployment, energy harvesting and transfer, and hardware solutions. The advanced physical layer techniques are surveyed in \cite{feng2013-survey} from an energy-saving perspective, and the impact of the physical layer design on energy savings is discussed. \cite{wang2012-survey} presents a comprehensive survey of green techniques for mobile networks with their associated advantages and limitations, covering various dimensions such as green metrics and energy performance trade-offs. 
\cite{muhammad2015-survey} highlights energy-efficient solutions (named ``Green solutions") and analytical models for energy consumption in base stations and mobile terminals. Based on traffic load conditions, solutions are identified and analyzed in this work. An extensive survey on ongoing research activities to improve energy efficiency for Heterogeneous Cloud Radio Access Networks (H-CRAN) and edge computing is presented in \cite{ali2018-survey} along with associated challenges and open issues. An evaluation of energy-saving solutions (based on research findings available) for Radio Access Network (RAN) is presented in \cite{larsen2023-survey} that showcases the reduction in energy consumption by approximately 30\% in mobile networks by choosing and implementing the most energy-efficient technology in RAN. \cite{hasan2011-survey} focuses on the role of heterogeneous network planning (considering options of micro, pico, and femtocells) in network energy saving. Further, it discusses cognitive radio and cooperative relaying aspects of green communication and introduces the concepts of energy-aware   cooperative Base station (BS) power management and energy-aware   medium access control. The focus of~\cite{ahmadi2025-survey} is to investigate Open RAN (ORAN)-based architecture from a sustainability perspective. It highlights the role of AI/ML and notarization in designing solutions for sustainability. The main center of interest for the above-mentioned survey papers is energy efficiency for radio access networks. However, these works do not cover design aspects related to energy awareness with AI/ML integration and support for information exposure required for energy-saving techniques in future networks, which is the primary focus of our survey.

\par Many papers also focus on the applications and energy-associated impacts of AI/ML integration in mobile networks from the energy efficiency perspective. A recent article~\cite{li2025-survey} on future 6G green wireless networks highlights the need for energy-efficient AI/ML-driven architecture designs and also details existing technologies used for energy efficiency. This work also proposes an intelligent architecture with an AI/ML model deployment framework to save energy and presents some preliminary results for energy saving. In~\cite{dbouk2025-survey}, AI/ML techniques are tabulated, which are being considered for saving energy at base stations in the 5G network.
\cite{you2025-survey} highlights concerns about high energy consumption due to large AI/ML models for achieving ubiquitous AI in 6G. A framework is proposed to integrate AI in sustainable 6G to address AI/ML-related aspects such as real-time controlling, and energy consumption to handle large data pools. Further, some lightweight AI/ML techniques are investigated from the perspective of energy efficiency. An approach of implementing slices from the energy-saving perspective in the mobile network is discussed in~\cite{lorincz2024-survey}, considering different energy needs of slicing use cases. The focus of these works is the applications and techniques associated with AI/ML integration in mobile networks. These papers are listed in Table \ref{survey} with their title and the domains covered.

Effective integration of AI/ML in networks cannot be achieved without the network supporting the following: (i)~collection and transportation of relevant data, (ii)~computation support for training and evaluation of AI/ML models, (iii)~delivery and updation of AI/ML models at appropriate network elements, and (iv) providing support for federated learning.
How this may impact architecture, interfaces and protocols in the existing mobile networks is not considered in the prior surveys. We cover this aspect in our survey.


\par There are some new technologies/approaches are being discussed for a sustainable future 
 mobile network. Symbiotic blockchain (SBC) 6G networks that are based on the convergence of Symbiotic communication and blockchain technology are surveyed in~\cite{luo2025-survey} from the perspective of energy saving in the network. A comparison of energy consumption is also provided based on simulation results using a sharding scheme in SBC networks. In another survey paper~\cite{pivoto}, relevance analysis methods are discussed for comparing 6G enabling technologies. Enable clusters are used to generate scores in different directions. Energy is one of the enable clusters; green technologies and energy harvesting are analyzed for energy enable clusters. A survey~\cite{singh2025-survey} reinvestigated the role of Software Defined Networking (SDN), Network Function Virtualization (NFV), and Cloud RAN (CRAN) from the perspective of enhancing energy efficiency in cellular networks. 
 The above-mentioned approaches look into the specific aspect of the network and do not cover the big picture from the perspective of 6G network design, which is the primary intent of our work.
 \begin{table*}[h]
 \centering
 \vspace{-0.3cm}
\caption{Abbreviations}
\begin{tabular}{|p{0.5cm} p{4.2cm}|p{1.1cm} p{4cm}| p{0.5cm} p{5.2cm}|}
\hline
 3GPP & Third Generation Partnership Project & EIF & Energy Information Function & NGMN & Next Generation Mobile Networks\\ \hline
 4G & Fourth Generation & FL & Federated Learning & OAM & Operations, Administration, and Management\\
 \hline
 5G & Fifth Generation & GSMA & GSM Association & OFDM & Orthogonal Frequency Division Multiplexing\\
 \hline 
6G & Sixth Generation & HAPS  & High Altitude Platform Station & ORAN & Open Radio Access Network\\
 \hline 
ADC & Analog to Digital Converter & H-CRAN & Heterogeneous CRAN  & QoE & Quality of Experience\\
 \hline 
AF & Application Function & ICT & Information and Communication Technology  & QoS & Quality of Service \\
 \hline 
AI & Artificial Intelligence & IEEE & Institute of Electrical and Electronics Engineers & RAT  & Radio Access Technology \\
 \hline 
AR & Augmented Reality & IMT & International Mobile Telecommunications & RAN  & Radio Access Network \\
 \hline 
BS & Base station & IRS & Intelligent Reflecting Surface & RF & Radio Frequency\\
 \hline 
CRAN & Cloud Radio Access Network & ITU & International Telecommunication Union & SBC & Symbiotic blockchain\\
 \hline 
CSI & Channel State Information & KPI & Key Performance Indicator & SDN & Software Defined Networking\\
 \hline
CSP & Communication Service Provider & LAN & Local Area Network & SG & Study Group\\
 \hline
DAC & Digital to Analog Converter & LEO & Low Earth Orbit & UAV & Unmanned Aerial Vehicle\\
 \hline
DL & Downlink & MIMO & Multi Input Multi Output  & UE & User Equipment\\
 \hline
DRL & Deep Reinforcement Learning & ML & Machine Learning & UL &  Uplink\\
 \hline
DRX & Discontinuous Reception & NEF & Network Exposure Function & UPF &  User Plane Function\\
 \hline
 DTX & Discontinuous Transmission & NFV & Network Function Virtualization &  &  \\
 \hline
\end{tabular}
\label{abb}
\end{table*}
 \begin{table*}[h!]
  \centering
  \vspace{-0.3cm}
  \caption{Energy-awareness in Mobile Network}
  \begin{tabular}{|p{1.8cm}|p{4.5cm}|p{10cm}|}
  \hline
    Literature & Energy related aspect & Highlights\\
     \hline 
\multicolumn{3}{|c|}{\textbf{Energy consumption in mobile networks}} \\ \hline \cite{ericssonblog,gsma2,williams2022, huttunen2023,singh2025,jorke2025} & Sources of energy consumption & Sources of energy consumption in mobile networks to highlight need for energy saving due to financial, environmental, operational, and societal reasons \\
     \hline \cite{hu2021energy, fowdur2023review, mao2021ai,kumar2023efficiency} & Advancements towards energy efficiency in 6G & AI/ML techniques to improve communication and decision-making, the integration of renewable energy sources, and the implementation of smart energy management systems to mitigate the anticipated rise in energy consumption \\
     \hline 
\multicolumn{3}{|c|}{\textbf{Energy efficiency as a QoS parameter}} \\ \hline 
\cite{theingi2024, pourkabirian2021} & Energy efficiency as a key metric for optimization & Energy efficiency maximized by optimum data transmission rate allocation for UEs or by optimizing resource allocation \\
\hline 
\cite{chou2024},\cite{hossfeld2024} & New energy metrics & ``Energy cost of AI life cycle (eCAL)" to estimate per bit AI/ML energy cost; ``Energy
intensity” defined as the energy consumed per data volume
in a time frame   \\
     \hline 
\multicolumn{3}{|c|}{\textbf{Usage of renewable energy sources}} \\ \hline 
\cite{greenlining2022, iitb1, ahmed2018, israr2024} & Prioritized deployment/utilization of Renewable energy sources & Cooperative micro-generation energy framework for operating base stations using renewable energy sources, potential benefits of using renewable energy sources
for IoT applications  \\
\hline 
\cite{Gelenbe, yuan2022, basaran2024} & Challenge associated with renewable energy sources & Varying characteristics
of renewable energy sources over time and usage of AI/ML for energy management for hybrid energy systems \\
     \hline 
\multicolumn{3}{|c|}{\textbf{Energy usage and energy saving trade-off by using AI/ML}} \\ \hline 
\cite{mao2022, gouissem2024} & AI/ML for energy efficiency  & AI/ML based learning approaches and algorithms to optimize energy efficiency \\
\hline 
\cite{chou2024, zhao+++2022} & Challenges associated  & Increased energy cost due to AI/ML and its evaluation \\
     \hline 
\cite{gouissem2024, zhao+++2022, chollet2023} & AI/ML energy trade-off &  Techniques for balancing AI/ML accuracy and its energy efficiency \\
     \hline 
     \multicolumn{3}{|c|}{\textbf{Green user satisfaction}} \\ \hline 
\cite{telco2022, capgemini} & User's choice and consent  & Service-wise energy awareness for environment-concerned users 
 \\
\hline 
\multicolumn{3}{|c|}{\textbf{Connectivity to energy constraint areas}} \\ \hline 
\cite{itu-report1, gsma2023-whitepaper} & Connectivity gaps  & Connectivity challenge 
to energy availability issues such as high energy costs, non-accessible renewable energy, and very frequent and prolonged
power outages  \\
     \hline 
\cite{ijala2022, kudu2022, shehab2022} & Few solutions &  Renewable energy-based deployments with energy harvesting, Digital smart grid mobile energy services\\
     \hline 
\multicolumn{3}{|c|}{\textbf{Energy saving techniques in the network}} \\ \hline 
\cite{ericssonblog, borja2024, lahdekorpi2017, laselva2024} & Dynamic configuration adaptation for hardware
components in RAN   & Conventional Discontinuous transmission
(DTX) / Discontinuous Reception (DRX) scheme  with multiple sleep states and considering multiple domains \\
     \hline 
\cite{younes2022, fall2023, ma2024, vallero2025} & Algorithms for energy saving &  Optimizing other parameters such as the spectral efficiency while minimizing the energy consumption and without compromising performance \\
     \hline 
 \cite{pan2022, zhang2025} & Energy harvesting &  Energy transfer techniques based on energy harvesting
architectures, Energy management considering hybrid energy models \\
     \hline 
 \cite{johansson2024, song2024, theingi2024} & Customized solutions &  Application specific solutions for cloud environment, High Altitude Platform Station (HAPS), Robotic airborne BSs  \\
     \hline      
 \cite{jorguseski2025, merluzzi2022, Duran2025} & Energy-aware   network decisions (new)  &  Network decisions considering energy
information (including energy usage and energy source type)  \\
     \hline  
\multicolumn{3}{|c|}{\textbf{Energy-aware   services/decisions support in the network}} \\ \hline 
\cite{tr22883, tr22882} & Energy information exposure   & Energy-related information exposure, e.g., energy consumption and energy source (renewable or non-renewable) at a granular level with awareness of energy source type \\
     \hline 
\cite{p1941} & Classification of energy-aware  service &  Service subscriptions for users with
multiple levels of energy and QoS mapping \\
     \hline 
 \cite{S2-2411073, S2-2411074, ts23501} & Energy information management in the network &  Coordination of service requirements and
the selection of network functions based on energy information (for example, energy usage, energy credits, etc.) \\
     \hline 
\cite{yan2019, lozano2025, azzino2024, zhao2025, lin2025, kolackova2025} & Energy models and parameters for mobile networks  &  Energy consumption model for core and RAN components for estimating end-to-end energy consumption for a service  \\
     \hline             
  \end{tabular}
  \vspace{-0.3cm}
    \label{literature}
\end{table*}

\subsection{Our Contributions}
This survey paper provides a comprehensive view of energy-aware  ness perspectives for existing and future mobile networks. The main contributions of this paper are as follows:
\begin{itemize}
\item We identify key motivations for considering energy awareness principles for designing future mobile networks. These motivation points are supported by research papers and industry reports. 
\item We propose design principles for energy-aware   mobile networks. These design principles are also investigated through ongoing works and available solutions from the perspective of feasibility in implementation. 
\item We present a few use cases for energy-aware   decisions to support the proposed design principles.
\item We discuss energy-aware   mobile networks' benefits from both the network and the user's perspective.
\item We explore and summarize standardization activities in this direction. We provide a list of ongoing items open for contributions in global standardization forums (Third Generation Partnership Project (3GPP), ITU, and Institute of Electrical and Electronics Engineers (IEEE)).
\item We list potential new research directions and challenges in this area.
\end{itemize}

The rest of the paper is organized as follows: Section~\ref{motivation}  motivates the need for including energy-aware   decisions and associated procedures in future network design. Section~\ref{design} describes the proposed design principles for energy-aware   mobile networks. Energy-aware network decisions are showcased with the help of use cases in Section~\ref{usecase}. Section~\ref{benefits} provides potential benefits of energy-aware   mobile networks from the network's and user's perspectives. Section~\ref{standards} covers standardization activities in 3GPP, ITU, and IEEE for energy-related aspects in mobile networks. Section~\ref{open} presents some thoughts on open research directions and challenges. Section~\ref{conclusion} concludes this work.  Table~\ref{abb} provides a list of abbreviations and Table~\ref{literature} presents a summary of references.

\section{Motivation towards Energy Awareness in the Mobile Network}
\label{motivation}
Energy awareness in mobile networks is driven by a range of multidimensional motivational factors aligned with sustainability goals on a broader level. This section highlights these key dimensions and respective accelerators for a sustainable network design.

\subsection{Energy Saving}
The primary source of energy consumption in wireless communication systems is RAN, particularly base stations. It is estimated that more than 70\% of the energy used by participating operators is consumed in the RAN, and within that, the base stations are the most significant energy consumers, with power amplifiers alone responsible for about 80\% of a base station's energy usage. The network core accounts for 10-15\%, owned data centers for 10\%, and other operations and support infrastructure for 5\% of the remaining energy \cite{ericssonblog,gsma2,williams2022, huttunen2023,singh2025,jorke2025}.
According to a 2021 GSM Association (GSMA) survey \cite{gsma2}, the primary energy efficiency ratio for RAN was 0.24 kWh per GB of data, with each mobile connection averaging 14.8 kWh annually and a typical network site using approximately 28,665 kWh over the same period.
If current trends persist, 6G could drive a 2–3X increase in network energy consumption compared to 5G, driven by denser deployments (e.g., small cells, mmWave), a surge in connected devices, and energy-intensive tasks such as real-time AI and digital twin applications.

Designing 6G networks with energy awareness is crucial for financial, environmental, operational, and societal reasons. As network density and data rates rise, energy costs could become unsustainable, potentially representing 40\% of a mobile operator’s expenses.
Environmentally, 6G networks risk significantly increasing the carbon footprint of the  Information and Communication Technology (ICT) sector, which accounts for nearly 3\% of global emissions. Energy-hungry systems may also limit deployment in areas with unstable power grids, hindering global connectivity goals. 
To this end, it is crucial to explore innovative solutions and technologies that can enhance energy efficiency in 6G, including advancements in network architecture, AI/ML techniques to improve communication and decision-making, the integration of renewable energy sources, and the implementation of smart energy management systems to mitigate the anticipated rise in energy consumption as we transition to 6G \cite{hu2021energy, fowdur2023review, mao2021ai,kumar2023efficiency}.
\subsection{Energy Efficiency as a QoS Parameter} 
Though energy efficiency is a Key Performance Indicator (KPI) at the network management level, there is no parameter directly related to energy as a part of Quality of Service (QoS) parameters. Besides, traditional metrics like bits per joule only captures energy consumption in data transmission but does not account for other energy consuming operations in the network like the energy cost of AI/ML integration-related components. 
Considering sustainability goals and AI/ML integration in 6G, the inclusion of an energy-related parameter in QoS can help the network and users to make energy-aware   intelligent decisions at the service level. There are many key considerations that will drive this energy parameter selection. A few such considerations are measurement feasibility based on available information exposure in the network, the inclusion of AI/ML energy cost, and the granularity of the energy information needed to evaluate that parameter. 
\par With this motive, summary of a few research works is as follows. An algorithm is proposed in \cite{theingi2024} for deploying and operating multiple Robotic Airborne Base Stations, which considers energy efficiency as a key metric by maintaining a balance between energy savings and ensuring maximal traffic loads. Energy efficiency is maximized by optimum data transmission rate allocation for User Equipments (UEs) based on accurate Channel State Information (CSI) obtained using maximum likelihood estimator \cite{pourkabirian2021}. A utility model is presented for energy exchange between a base station, electric vehicles, and grid, and further for edge computing task offloading from the base station to EVs for efficient resource utilization\cite{ayaz2024}. In other work, a new metric called energy cost of AI life cycle (eCAL) is proposed to estimate per bit AI/ML energy cost for an energy balanced decision making for optimum usage of AI/ML\cite{chou2024}. Another metric named ``energy intensity" is defined as the energy consumed per data volume in a time frame, and it is investigated in comparison with traditional energy efficiency parameters\cite{hossfeld2024}. These above-mentioned new parameters are defined for specific application scenarios and have limited scope. There is a need to rethink a common and feasible energy parameter that can be included in QoS parameters from a broader perspective.

\subsection{Usage of Renewable Energy Sources}
Integrating renewable energy into future 6G networks is essential for achieving sustainable, low-carbon operations and meeting the substantial power demands of ultra-dense, high-performance infrastructures by leveraging innovations in network architecture, seamless integration of renewable sources, and intelligent energy management systems to curb escalating energy consumption.
 The increased and prioritized usage of renewable energy sources helps in reducing the carbon footprint of future mobile networks. From many perspectives such as environment, social and energy cost, renewable energy-based deployments greatly increase energy affordability for connectivity and decarbonization in future 6G networks~\cite{greenlining2022}.
\par There are different ways to increase the use of renewable energy sources in mobile networks. Some are as follows: An energy-aware   service delivery approach can facilitate prioritization of renewable energy source-operated network resources over conventional energy-operated network resources while providing a service~\cite{iitb1}. For example, a user can receive a service via a Wireless Local Area Network (LAN) access point (say operated by renewable energy) instead of a 5G New Radio base station (say, operated by conventional energy). The feasibility of operating base stations using renewable energy sources is investigated in~\cite{ahmed2018}. However, cooperation between base stations is required, including renewable energy harvesting and cooperative micro-generation energy frameworks, as suggested in~\cite{israr2024}. The potential usage of renewable energy sources for IoT applications is detailed in~\cite{rahman2022} that showcases results in terms of energy cost and system robustness.
\par It is also to be noted that, due to the varying characteristics of renewable energy sources over time and space, QoS may need to be adjusted to enable service delivery while using renewable energy sources. In this context, the modeling of renewable energy sources and mobile traffic as random processes is presented in \cite{Gelenbe}. Further AI/ML-based learning can facilitate the regulation of dynamic attributes of renewable energy sources in mobile networks \cite{yuan2022}. An AI/ML based energy management concept is proposed to prolong the operating lifetime of base stations using renewable energy sources when they are suffering from the outage of conventional energy sources after natural disasters \cite{basaran2024}.  To summarize, solutions need to be developed to prioritize the usage of renewable energy sources over the grid power supply as part of the 6G design, considering the challenges associated with such energy sources.
\subsection{Energy Usage and Energy Saving Trade-off for AI/ML}
AI/ML-based optimization has been explored as a promising solution for achieving high energy efficiency in mobile networks. Numerous heuristic, traditional optimization techniques and deep learning AI/ML algorithms with different learning approaches have been utilized to optimize energy efficiency~\cite{mao2022}. 
{Trends in AI/ML algorithms for energy efficiency are detailed in \cite{mahmood2022} specifically from the perspective of supporting IoT applications in 6G. Highlighted trends include heuristic algorithms, supervised learning, unsupervised learning, reinforcement learning, deep learning, deep reinforcement learning and Federated Learning (FL). Similar trends and respective algorithms are presented in another review for various network operations for example resource allocation, task offloading, and handover management etc, considering a range of diverse applications including emerging networks like vehicular network, and Fog-radio access networks\cite{noman2023}.} However, with the expanded heterogeneity and massive hybrid usage in 6G networks, there is growing concern about the trade-off between achieving energy efficiency in 6G and the energy consumption of AI/ML algorithms.

{Research work in \cite{chou2024} highlights increased energy costs due to AI/ML integration in the system from the perspective of 6G networks. Further, a new metric named ``energy cost of AI life cycle (eCAL)" is proposed for the balanced evaluation of such systems. FL methods are explored to enhance energy efficiency in mobile networks in a survey~\cite{gouissem2024}.}
Some challenges associated with energy-efficient FL is highlighted, such as heterogeneity of devices, balancing between the accuracy of the model and its energy efficiency, and energy consumption due to AI/ML-related data exchange in the network. Similar challenges associated with FL-based learning in 6G are presented in~\cite{zhao+++2022}, and energy models for a few FL schemes are analyzed to compare energy consumption. To avoid AI/ML energy constraints for low-power devices, computational offloading is proposed in ~\cite{chollet2023}. Hence, considering the energy-AI/ML usage trade-off is essential when designing intelligent sustainable 6G networks.
\subsection{Green User Satisfaction}
User choice is key to encouraging Communication Service Providers' (CSPs) efforts toward sustainable solutions in the network~\cite{telco2022}. If energy awareness at the service level is provided to users, they can make a well-informed selection of services. Environment-concerned users (say, Green users) may choose to sacrifice QoS for energy efficiency. This choice also depends on the type of application. Thus, energy-aware   decisions in the network must be a coordinated approach that takes into account user preferences and application requirements. Hence, user consent is necessary for such approaches to ensure user satisfaction. In this context, applications and tools are being developed to analyze user usage trends and the associated impact on carbon footprints. Users can be incentivized by making energy-aware   decisions. For example, Capgemini, Nokia, and Google have developed a product called ``Energy efficiency for Scope3 indirect emissions" in collaboration \cite{capgemini}. It is an on-device app that can provide energy scores for the user based on their usage data. Such efforts by CSPs can lead to energy-saving solutions. However, the impact of these solutions needs to be discussed from the perspective of user satisfaction and trust, which is the final goal for CSPs.  
\subsection{Connectivity to Energy Constraint Areas} 
Although “connectivity everywhere and for everyone” has
been a continuous effort of the mobile networks’ community,
the number of people not connected to the internet is estimated to
be 2.6 billion worldwide as per the recent ITU press release \cite{itu-report1}. Energy availability is one of the barriers to providing connectivity in energy-constrained areas. As per a GSMA study report \cite{gsma2023-whitepaper}, Subsaharan Africa has a big connectivity gap as 76\% of its population does not have mobile internet subscriptions yet (data is for the end of the year 2022) where this number is 38\% for rest of the world. This report indicates that the connectivity challenge in Subsaharan Africa is due to energy availability issues such as high energy costs, non-accessible renewable energy, and very frequent and prolonged power outages.  
\par One possible solution is renewable energy-based deployments with energy harvesting\cite{ijala2022}. However, it is challenging due to the random behavior of renewable energy sources, which depend on environmental conditions. A concept of ``Digital smart grid mobile energy services" that supports deploying the intelligent multi-agent monitoring system, expanding renewable energy services, and regulating energy services is proposed in~\cite{kudu2022}. The issues related to these concepts are highlighted as follows: security, cost of operation, and acceptability of power energy sectors. In this context, the role of mobile networks in building sustainable and energy-efficient smart solutions for providing meaningful connectivity is presented in~\cite{shehab2022}. Further, the importance of low-energy network design and customized local solutions to reduce energy consumption in energy constraint areas for extending connected regions needs to be explored.


\section{Design Principles for Energy-aware   Mobile Networks}\label{design}
\subsection{Energy Saving Techniques in the Network} Energy saving is a key objective for future mobile networks, and new energy-saving solutions can coexist with techniques/solutions that are already being exercised in the existing networks. In this context, a few of the conventional methods and research trends are as follows.
\subsubsection{Dynamic Configuration Adaptation for Hardware Components in RAN}
The conventional Discontinuous transmission (DTX)\,/\,Discontinuous Reception (DRX) scheme for switching on/off Base Stations (BSs) is now advanced with a new concept for deciding sleep states (deep sleep, microsleep, light sleep, active Uplink (UL), active Downlink (DL)) by considering different time cycles and load conditions. The classification of sleep states is primarily based on transition times of the network components~\cite{ericssonblog}. In micro sleep state, components with the fast transition time (in few microseconds) are switched off by the network. Typically, amplifiers and chip-based small hardware components are included in this category. In light sleep, components having medium transition time (in the range of several milliseconds) are turned off in addition to fast transition time components. In a deep sleep state, many additional components having a transition time of several tens of milliseconds are switched off. However, it takes time to come back from this state to the ON state. Based on components being turned off in each state, the power level is highest in a micro-sleep state and lowest in a deep sleep state. The aim of this scheme is to achieve power saving using the DTX/DRX scheme without compromising Quality of Experience (QoE)~\cite{ericssonblog, borja2024,lahdekorpi2017}. Multiple domain adaption in radio across time, frequency, antenna, and power domains is compared with single domain adaptation in~\cite{laselva2024}. For this technique, implementation of dynamic configuration for hardware components is required based on real-time requirements.

\subsubsection{Algorithms to Optimize Energy Efficiency} 
Striking a better trade-off between energy efficiency and performance is modeled as an optimization problem in many research works. A few of them are discussed in this subsection. The trade-off between spectral efficiency and energy consumption is verified, and a solution for optimizing the spectral efficiency while minimizing the energy consumption is proposed in~\cite{younes2022}. An energy consumption optimization algorithm for cell selection between macro and pico cells with consideration of high-performance requirements is proposed in~\cite{fall2023}. A proactive data-driven energy-saving method named REDEEM, which deploys strategic 5G traffic offloading to overlapping Fourth Generation (4G) cells based on data predictions, is verified through experiments in~\cite{ma2024}. This method can be made applicable to offloading traffic between multiple Radio Access Technologies (RATs) in heterogeneous networks. An approach of ``decreasing transmitted power until possible" is investigated in \cite{vallero2025} considering a threshold for transmission power for active BSs while ensuring coverage for users. With advancements in AI/ML, learning-based energy optimization algorithms are now being considered.

\subsubsection{Energy Harvesting}
{Efficient energy transfer techniques assure optimum energy usage and management in the network. Energy harvesting is such a technique in which low-power devices like sensors, relay nodes, and even mobile handsets capture and convert ambient energy from the environment into usable electrical power.} Hence, energy transfer techniques based on energy harvesting architectures are considered as feasible solutions for achieving sustainability goals. These solutions are especially more rewarding for increasing the lifetime of low-energy devices such as IoT devices. However, some concerns related to the security and effectiveness of energy transfer using such solutions exist. Techniques like differential privacy are being used for making secure energy transfer \cite{pan2022}. A solution of using Intelligent Reflecting Surface (IRS)-aided energy transfer is proposed in \cite{pan2022} to increase energy transfer efficiency. IRS is used to adjust the amplitude and phase (beam forming) of energy signals to maximize the energy transfer. In hybrid energy models, uneven distribution of energy across various nodes results in higher grid energy utilization. In a research work, a grid energy consumption optimization problem is solved using an algorithm considering a joint strategy for power control and energy management \cite{zhang2025}. With privacy and intelligent solutions, energy harvesting is one of the most important features of 6G networks, especially for low-power devices.

\subsubsection{Customized Solutions}
Some of the above-mentioned techniques are based on sleep modes or algorithms, which can be considered as part of network design principles and can be included as part of 6G network design principles. However, few techniques are application or implementation-specific, which are not part of the scope of network design. For example, the configuration of the shim layer is proposed as an energy-saving mechanism in the cloud environment \cite{johansson2024}. Application-specific methods such as core frequency scaling and uncore frequency scaling are also proposed for cloud environments. High Altitude Platform Station (HAPS) based base stations are proven to be more energy efficient as compared to terrestrial base stations due to their unique advantages like wide coverage and self-sustainability \cite{song2024}. This is a deployment-specific method used to achieve high energy efficiency. Robotic airborne base stations are also proposed and analyzed from the energy efficiency perspective in \cite{theingi2024}.
\subsubsection{Energy-Aware   Network Decisions}
Besides the above-mentioned techniques, making energy-aware   network decisions is a new approach for increasing energy savings. For example, energy information (including energy usage and energy source type) can be taken into account in resource allocation mechanisms. Such energy-aware   resource selection decisions can result in minimizing energy usage and increasing the renewable energy usage ratio in the network to a great extent. Making energy-aware   network decisions is possible only with enhanced exposure to network energy information at a granular level (per user, slice, network element). However, there is a limited discussion and ongoing work in this direction for mobile networks. From a broader perspective, enablers for energy-aware   Information and Communications Technology (ICT) systems are presented in~\cite{jorguseski2025}. From another point of view, energy awareness-related goals are being considered in AI/ML-based solutions. In a related work~\cite{merluzzi2022}, energy consumption is regarded as a cost for accomplishing inference effectiveness (accuracy and reliability), and the aim is to minimize energy cost subject to achieving the desired effectiveness. An energy-aware   reward function is used in the Deep Deterministic Policy Gradient algorithm-based learner model for 6G IoT systems~\cite{Duran2025}. Further, the approach of including energy awareness in network decisions needs to be investigated from the perspective of its inclusion as a design principle for 6G networks. 
\subsection{Energy Information Exposure} 
The exchange of energy-related information has been part of procedures and performance analysis in mobile networks. However, energy-related information exposure, e.g., energy consumption and energy source (renewable or non-renewable)  at a granular level with awareness of energy source type, is a new direction now being considered. In this context, exposure of energy-related information to verticals and users is being considered in recent releases of 3GPP standards \cite{tr22883}, \cite{tr22882}. Even though networks can utilize energy-related information to support energy-aware   service, exposure of energy-related information to users and the consent of users for energy-aware   service delivery may be required, as it may result in dynamic adjustment to QoS. The existing 5G system supports energy consumption monitoring per network slice and subscriber level. The monitoring could be based on a statistical model of energy consumption and not necessarily done in real-time. As per 3GPP specifications, the 5GS shall support monitoring energy consumption for serving a 3rd party, which may relate to energy consumption by network resources of a network slice, a Non-Public Network (NPN), etc. Further proposal is to include information on carbon emissions or renewable energy ratio and energy credit limit, e.g., the information about the difference between the consumed energy and energy credit limit for the services to a subscriber. It shall also support the exposure of performance statistics information of the network, e.g., the data rate, the packet delay, and the packet loss, along with the energy consumption information to external entities. This information can help correlate energy consumption with the QoS supplied. 
\par Based on ongoing discussion for existing mobile networks, to facilitate energy-aware   service delivery, support for energy consumption monitoring at a granular level, i.e., per service, per flow, per user along with the ability to identify energy source(s) used (including information on renewable and non-renewable energy sources used along with their ratio in the mix) in the 6G network is proposed, as illustrated in Figure 1. Once collected (through Operations, Administration, and Management (OAM)), the network can provide users with service-level energy usage information, including information on energy sources. The energy usage exposure facilitates the provisioning of energy-aware   services in the network based on user choice. 
 \begin{figure}[ht]
\vspace{-0.2cm}
\centering
\includegraphics[width=\columnwidth]{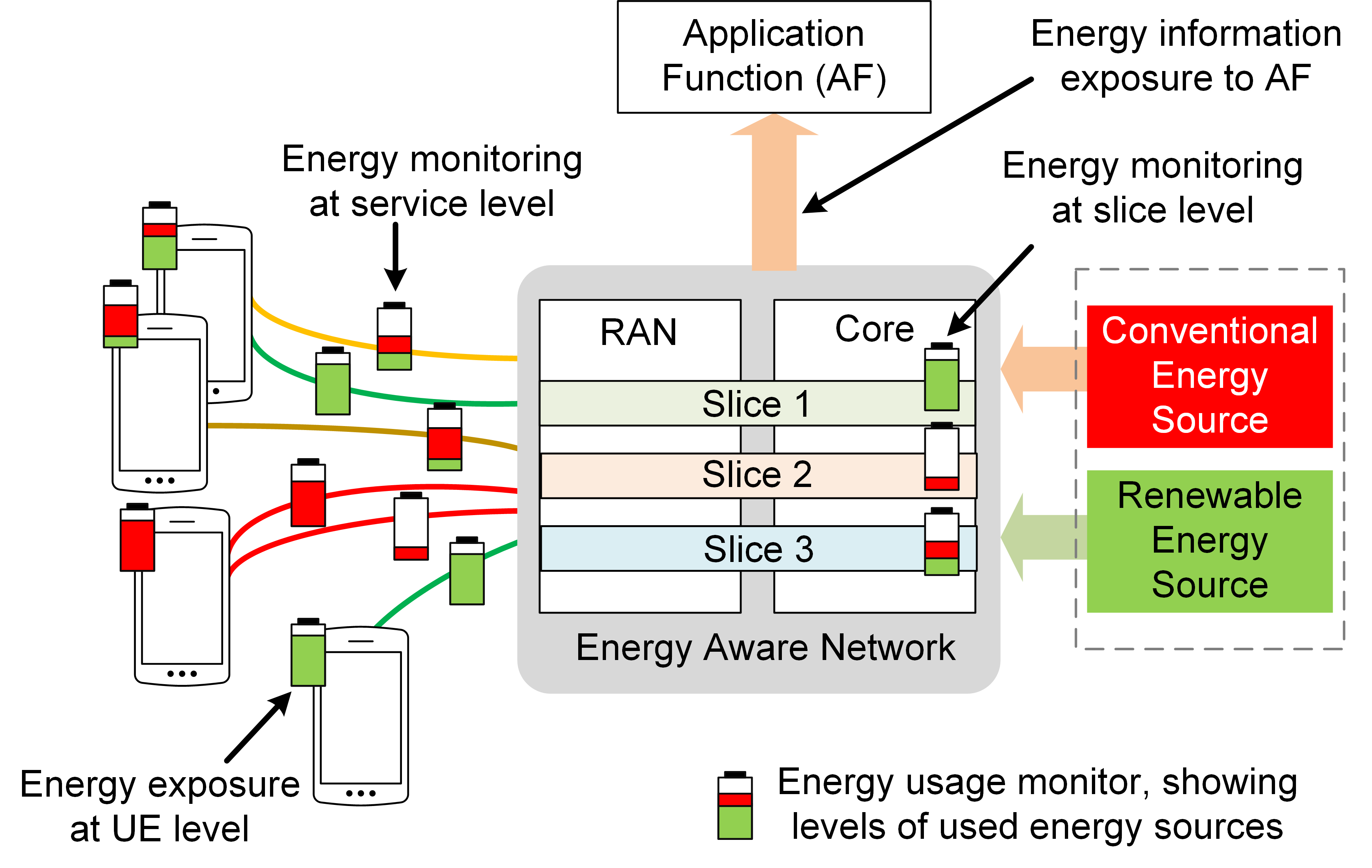}
\vspace{-0.4cm}
\caption{Energy monitoring at granular level in Energy-aware   6G network.}
\label{ce}
\vspace{-0.2cm}
\end{figure}
\subsection{Classification of Energy-aware   Services} 
\begin{figure}[ht]
\vspace{-0.2cm}
\centering
\includegraphics[width=\columnwidth]{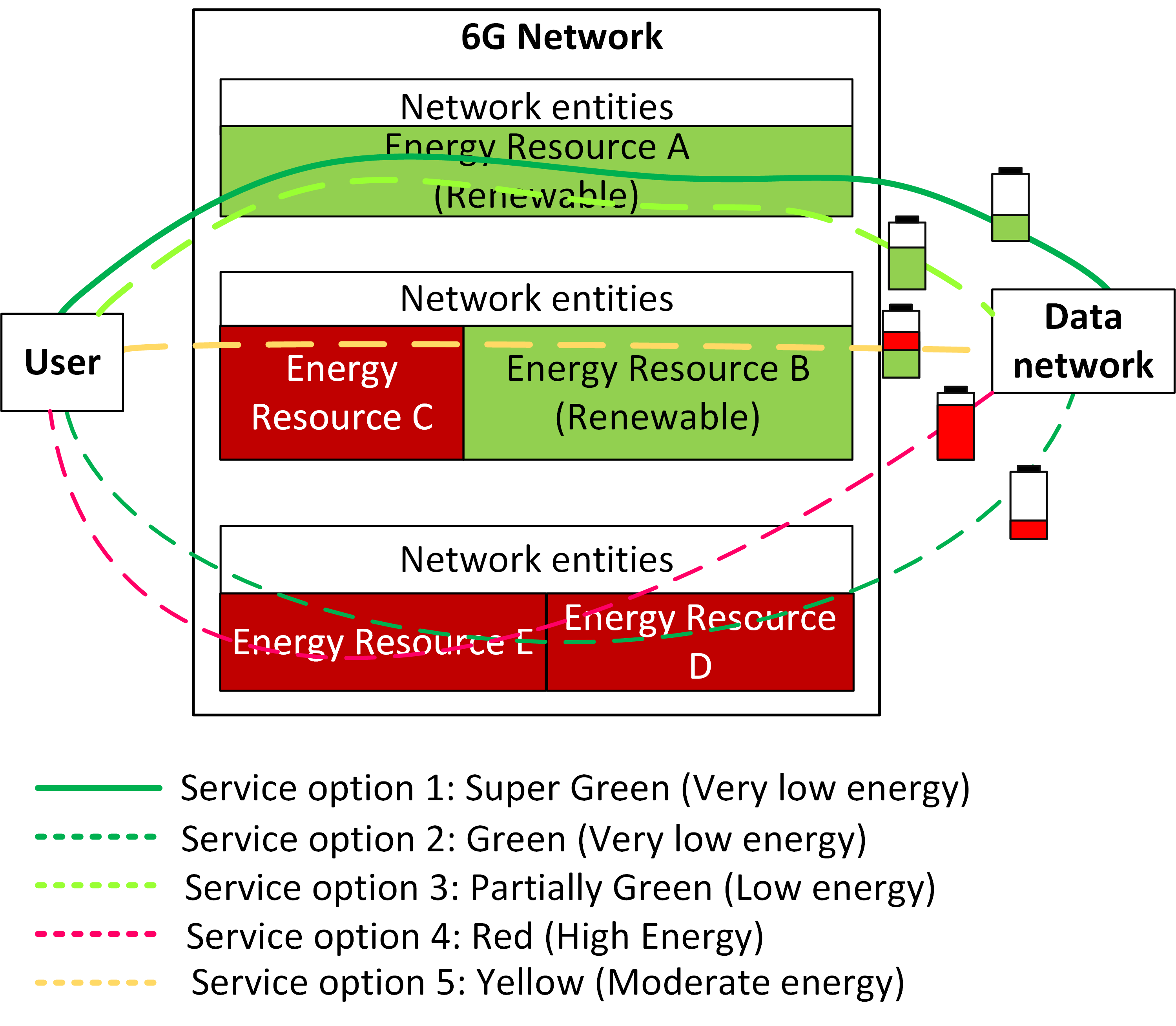}
\vspace{-0.6cm}
\caption{Some examples of Energy-aware   services.}
\label{es}
\vspace{-0.2cm}
\end{figure}

A service provider can support “energy-aware   services”, which can be classified based on energy information with associated QoS. The user can subscribe to the energy-aware   service when a user wants a service that considers energy usage and the type of energy sources used in mobile networks. By subscribing to such a service, the user has consented to dynamic service quality adjustments based on energy-aware   network decisions. It can be the choice of users/verticals to select an option from available energy-aware   services as illustrated in Figure \ref{es}. For example, an environment-friendly user chooses a super green service (very low energy, very low QoS, provided by renewable energy sources only) over a red service (high energy usage, high QoS, provided by conventional energy sources). Table \ref{est} classifies different types of services based on energy usage and associated QoS values. Please note that these are only some examples of energy-aware   services and do not cover all the possible options.
\par The feasibility of service subscriptions for users with multiple levels of energy and QoS mapping can introduce openness in energy-based network exposure. QoS comprises of many parameters, some of which are as follows: bit rate, packet delay budget, and packet loss ratio. A scheme can be used to convert these QoS values into QoS levels, and further can be directly mapped to energy consumption. This can also be made related to pricing policies, so low-energy users can also avail themselves of cost benefits. For example, a high data rate can lead to more energy usage in the network than a low data rate. Additionally, including energy consumption as a performance criterion, like bit rate and latency, is the first step towards designing energy-aware   services and subscriptions.
\begin{table}[h!]
\caption{Examples for classification of Energy-aware   services}
\label{est}
\centering
\begin{tabular}{|p{0.28\columnwidth} | p{0.19\columnwidth}| p{0.21\columnwidth} | p{0.12\columnwidth}|}
\hline
\textbf{Service classification} & \textbf{Energy usage} & \textbf{Energy source}* & \textbf{QoS}\\ \hline
Super Green & Very low & R only & Very low\\ \hline
Green & Very low & C only & Very low
\\\hline
Partially Green & low & R only & Low
\\\hline
Yellow & Moderate & R and C both & Moderate
\\ \hline
Red & High & C only & High
\\\hline
\end{tabular}
*R: Renewable energy source, C: Conventional energy source
\end{table}
\subsection{Energy Information Management in the Network}
A dedicated Energy Information Function (EIF) is recently been introduced to coordinate service requirements and the selection of network functions based on energy information (for example, energy usage, energy credits, etc.) in 3GPP SA2 working group \cite{S2-2411073}, \cite{S2-2411074}, \cite{ts23501}. Figure~\ref{ce} shows the collection and exposure of energy information in the network through EIF. EIF collects granular-level energy information from OAM and other real-time information from all network functions to analyze and monitor energy parameters at a granular level (per user/ per service). Application Function (AF) shares service requirements with policy control function during service setup. EIF also interacts with session management functions in the core and provides information regarding service (flow) wise energy information. Session management functions can utilize this information for energy-aware   resource allocation, which further may need adjustments in QoS at the flow level. By “Energy aware”, we mean to have energy information, considering this information and energy usage constraint for all associated activities.
\begin{figure}[ht]
\vspace{-0.4cm}
\centering
\includegraphics[width=\columnwidth]{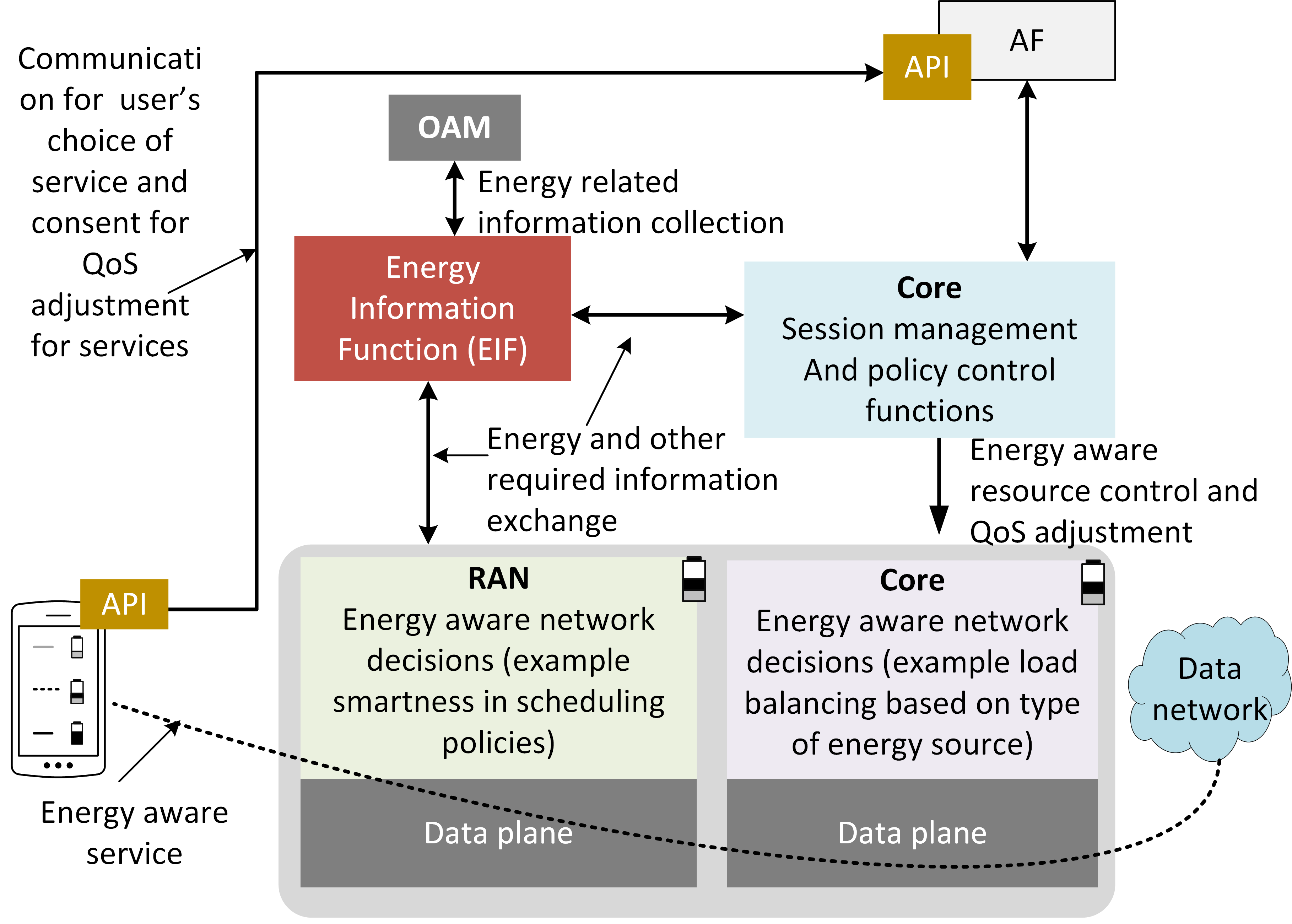}
\vspace{-0.4cm}
\caption{Collection and exposure of energy information in the network through a new energy function EIF.}
\label{ce}
\vspace{-0.2cm}
\end{figure}
\par The energy information exposure within the network can drive energy-aware   decisions in Radio Access Network (RAN) and core, and these decisions at various levels adjust policies and QoS per flow/service for underlying resources. One example of energy-aware   decision is energy-aware   intelligent scheduling in RAN, which can apply changes in end-to-end delay to achieve higher energy efficiency. If the delay budget for a user is increased, the base station has greater flexibility in scheduling the user. It can decide to schedule the user when the radio condition is better for data transfer to that user, meaning more energy efficiency can be achieved. Increased buffering capacity, e.g. playback buffer capacity for video application, may also help in energy-efficient operation, allowing for energy-efficient scheduling. Another example is energy-aware   load balancing-related decisions in the core. The core can prioritize using renewable energy powered network elements to manage the maximum possible load. EIF also maps energy information at the service level, along with associated QoS, and this mapping can be exposed to AF as per their choices. Overall, a dedicated function (EIF) for managing energy-related information and coordinating energy-related communication is a must for a feasible energy-aware network. 
\subsection{QoS Adjustment Based on Energy Information in the Network}
Energy-aware network decisions and provisioning of energy-aware   services may result in QoS adjustments. However, user consent is mandatory for such energy awareness-based policies and is attained during registration of services. A few example scenarios to trigger QoS adjustments are dynamic availability of renewable energy sources, energy credit limit of the user, and energy usage constraints of resources. For example, suppose there is limited energy availability at a renewable energy powered BS, then transmission policy, and thereby QoS of services, can be adjusted to save energy.  
Another example scenario can be based on the type of service, where energy-aware   QoS adjustment for services can be applied. Consider a user is watching a video over the mobile network. To reduce energy usage (consumption) in video delivery, the service provider can reduce the resolution from high-definition video to standard-definition video. The resultant decrease in the data rate reduces the energy consumption in the network. However, QoS adjustment can’t be made for all types of services, as a voice call may get interrupted with QoS adjustment. Therefore, the network has a configurable range of acceptable QoS requirements for energy-aware   services. Accordingly, dynamic QoS adjustment (within an acceptable pre-defined range based on service type) is applied at the flow level. It is a trade-off between energy consumption and QoS, but it is the user’s choice with support from the network.



\par Network has information about the QoS of individual flows. Using the Application Function (AF)/Network Exposure Function (NEF) interface for NEF-AF session (with QoS service) or NAS level PDU Session related messages from UE, the network can acquire the QoS-related information for individual flows (QoS rules, QoS flow descriptions). AF can provide alternative service requirements for each service (flow) containing QoS parameters in a prioritized order to be selected dynamically by the network. Based on inputs from UE (regarding user’s consent on QoS adjustment) and the energy information collected from the network functions through EIF, the policy control function can make energy-aware   policy decisions to adjust QoS dynamically for services (flows). This is how energy-aware   QoS adjustment can be supported in 6G mobile networks.

\subsection{Energy Models and Parameters for Mobile Networks}
Considering energy-aware  ness at user and service level, service-wise energy consumption estimation is a crucial aspect for making energy-aware   network decisions. Based on the service type, different topologies can be considered for end-to-end energy estimation. General topologies for the mobile network services are user-to-data network (e.g., web browsing), user-to-user direct (e.g., phone call), or via data network ( e.g., Virtual Reality/ Augmented Reality (AR)). Interdependency between load-related parameters at different network nodes also needs to be taken care of while modeling end-to-end energy consumption for a service. Based on prior art available, the following are the energy models being considered for core and radio access network (note that energy consumption modeling for data network and user equipment is out of the scope of this paper):
\subsubsection{Energy Consumption Model for Core Network Elements} A minimal discussion is available for energy modeling of core network elements. In  \cite{yan2019}, a parameter $E_{core}$ representing {service-specific} energy estimation for core is evaluated as follows: 
\begin{equation*}
E_{core} = (N_c E_c + N_e E_e + E_{g} + E_s) \int_0^{T_d} D_d \, dt,
\end{equation*}
where $N_c$ and $E_c$ denote the number of core routers and energy per bit for core router respectively; $N_e$ and $E_e$ denote the number of edge routers and energy per bit for edge router respectively; $E_g$ denotes energy per bit for gateway; $E_s$ denotes energy per bit for switches; $T_d$ denotes time duration for data; and $D_t$ denotes data traffic in bits per second. It is important to note that the above expression only considers energy associated with data transmission. However, energy for signaling, which is required for session establishment before data transmission, is not covered.
\subsubsection{Energy Consumption Model for Radio Access Node (Base Station)} There are many directions covered in available literature for energy modeling of radio access nodes in mobile networks. Few of them are discussed in this section. As per \cite{yan2019}, base station energy consumption $E_{bs}$ can be derived as follows:
\begin{equation*}
E_{bs}=  \int_0^{T_d} P_{bs(d)} \, dt + \int_0^{T_s} P_{bs(s)} \, dt.
\end{equation*}
Here $T_d$ and $T_s$ represent time duration for data and signaling, respectively; and $P_{bs(d)}$ and $P_{bs(s)}$ represent power consumption of base station in watts for data and signaling transmissions, respectively. Further, $P_{bs}$ has static and load dependent components \cite{azzino2024}, \cite{zhao2025}, \cite{lozano2025}  as follows:
\begin{equation*}
P_{bs}(i,l) = P_S(i) + L P_D(i,l),
\end{equation*}
where $P_S(i)$ denotes static power consumption and $P_D(i,l)$ denotes load ($l$) dependent dynamic power consumption for $i^{th}$ base station; and $L$ represents load coefficient. Static power consumption for a base station $P_S(i)$ consists of power consumption of the baseline components in radio unit (e.g., interfaces), Radio Frequency (RF) components like Analog to Digital Converter (ADC)/ Digital to Analog Converter (DAC), and frequency up/down converters and core digital processors \cite{lozano2025}. And, dynamic power consumption is impacted by power amplifiers and their efficiencies in varying load conditions.
\par However, energy modeling for network elements has been worked upon while energy efficiency optimization techniques have been applied. Its scope has focused on energy models for RAT nodes (Base stations). Service type can impact the overall energy consumption estimation, and service probability can be used to generate a probabilistic model for service-wise end-to-end energy consumption. Other than energy consumption, energy efficiency (Energy per traffic load) and energy saving rate (considering sleep modes) are the parameters being explored \cite{lin2025}, \cite{kolackova2025}. However, there is no standardized way for energy modeling and its associated parameters to include energy awareness into 6G network design and its procedures.
\subsection{Joint Optimization of Network Performance, Resources, and Energy Using AI/ML}

The incorporation of AI/ML into the network is touted as a key technology for the 6G system, alongside improved energy awareness and standardized energy models. An `AI Native' 6G architecture, characterized by learning-based network design and algorithms, is expected to substantially enhance resource allocation mechanisms and the overall 6G system \cite{8808168,9446676}. To this end, there is a considerable difference in the energy consumption between the integration of AI into the network (or native AI) and use of external AI support (Patch-on AI) \cite{cui2023}. The Rel-18 of 3GPP emphasises the adoption of learning-based optimization for the New Radio (NR) Air Interface, especially for energy savings, resource allocation, load balancing, mobility, and interference management \cite{9795045}. Consequently, AI facilitates the objective of \textit{joint optimization} by integrating `energy awareness into design and decision-making' alongside `network performance optimization'.

 To this end, the energy usage of network resources should be taken into account when deciding algorithms' goals along with other QoS requirements.
 {An important challenge is to model the energy consumption of diverse network functions accurately, as this is critical for enabling effective joint optimization.} Several AI/ML-based approaches have been investigated for optimizing a combined objective comprising of network performance, compute and energy utilization, and selection of energy source. We will broadly organize them on the basis of their underlying ML methodologies.

\subsubsection{Supervised Learning}
{Supervised learning is an ML approach in which the algorithm is trained on a labeled data set, which means that each input is paired with a known output, allowing the model to learn the relationship between inputs and outputs and make accurate predictions on new, unseen data. 
Several ML models have been developed for the joint optimization of network performance, resource allocation, and energy efficiency in wireless systems. Here, the input data typically includes metrics such as channel state information, traffic and device mobility patterns, and energy consumption profiles, which are used to train models that make real-time optimization decisions.}
A scheduled energy harvesting in wireless power transfer using linear regression is studied in \cite{9121977}. A classification framework for base stations, based on their spectral and energy efficiency, was introduced in \cite{mata2025}, which uses supervised learning to determine the influencing factors. The authors in \cite{8741057} investigated the application of multiple ML techniques for the optimal prediction of green energy, brown energy, and QoS combination in RAN.

\subsubsection{Unsupervised Learning} Unsupervised learning is a learning mechanism where input data is directly used to learn the underlying pattern without a labeled training set. This method relies on identifying underlying pattern of the input data to perform optimization decisions on real-time data. A Deep Neural Network (DNN) using online and unsupervised learning methods for energy-efficient OFDM transmission is studied in~\cite{10283910}. Small cell and Heterogeneous Networks (HetNet) are critical techniques for improved performance in wireless networks, and a multilayer HetNet-based strategy for joint optimization considering AI/ML energy consumption is discussed in~\cite{9839644}. A primal-dual unsupervised learning based DNN is implemented in \cite{9794291} to enable energy efficient power control methods over fading channels.

\subsubsection{Reinforcement Learning} Reinforcement Learning (RL) is a reward/penalty oriented learning where agents learn the optimal decisions through trial and error. The system to be learned is called an environment and the agent learns the optimal decision through a series of interactions with the environment with reward maximization/penalty minimization objective. A Deep Reinforcement Learning (DRL) based approach for energy saving at the cell level by cell on-off and antenna muting was considered in \cite{bassoy2023}. A hierarchical and distributed resource management framework was proposed in \cite{lee2022} to utilize DRL and achieve efficient radio resource allocation while prioritizing renewable energy sources. Meta RL to solve a sustainable RAN slicing scheduler problem for the joint optimization of resource allocation and energy consumption was studied in \cite{you2023}.  \cite{MUGHEES2023102206} presents a Multi-Agent Parameterised Deep Reinforcement Learning (MA-PDRL) approach for joint optimization of energy utilization and resource allocation in the HetNet paradigm. 

\subsubsection{Federated Learning} Federated learning is a distributed paradigm in which multiple devices collaboratively train a shared model while keeping data local and exchanging only model parameters. 
It is particularly useful when individual devices lack sufficient data to train an effective model, while sharing raw data is restricted due to privacy concerns.
A federated learning-based integrated sensing and communications system utilizing the same radio frequency in Massive MIMO for 6G networks is proposed in \cite{10620738}. Low Earth Orbit (LEO) satellites are a promising direction to boost coverage and efficiency in 6G communication. Decentralized Satellite Federated Learning (DSFL) is proposed in~\cite{10092560} for energy-aware   offloading of training task-specific machine learning models to satellites and enabling global access for the same.

Most energy-aware communication studies focus on emerging 6G technologies, including Intelligent Reflecting Surfaces (IRS), UAV-assisted connectivity, and application domains such as IoT and wireless sensor networks. While there has been some effort towards the joint optimization of resources and energy usage in 6G networks, there is still a considerable opportunity, especially in the 6G core network. The next generation wireless network is believed to support numerous applications enhanced by novel techniques, necessitating use-case specific research focus. This has led to a considerable effort towards more application oriented research as opposed to the core network. With increasing focus on environmental impact of modern communications, it is pertinent to focus on efficient and green energy-aware  design of the RAN as in \cite{10772596}. Through this work we intend to provide some energy-aware   AI-driven conceptual solutions for the combined optimization of resource allocation, load balancing, adaptive security, mobility and interference management while being conscious of the energy consumption of the AI solution itself.

\section{Examples of Adaptive Network Decisions based on Energy-aware   Design Principles}\label{usecase}
\subsection{ML-Based Energy-Aware Resource Allocation in the Core Network with Guaranteed QoS}
\begin{figure}[ht]
\vspace{-0.2cm}
\centering
\includegraphics[width=\columnwidth]{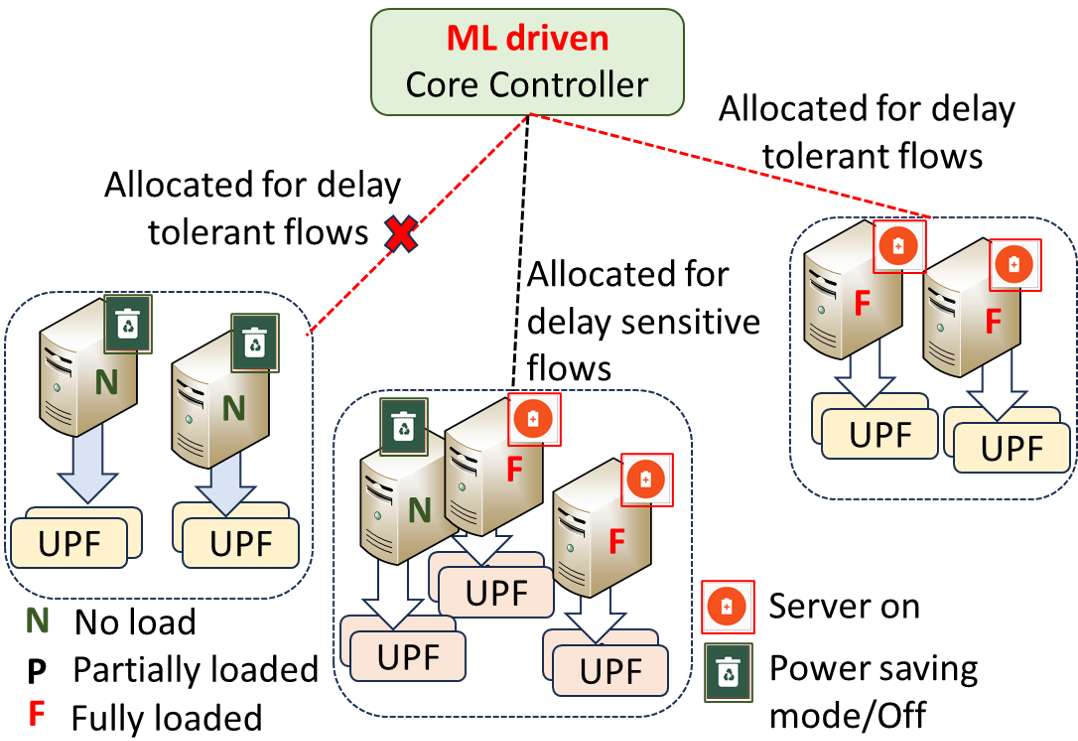}
\vspace{-0.4cm}
\caption{Energy-aware ML based resource allocation in core assuring required QoS.}
\label{upf}
\vspace{-0.2cm}
\end{figure}

A key use case involves integrating energy awareness into a real-time resource allocation, potentially adopting strategies, such as minimizing the number of active core network functions and putting others in `sleep' mode, to achieve energy savings in the network. Figure \ref{upf} shows an example of utilizing User Plane Function (UPF) based on energy information and QoS requirements to its full capacity (shown in the figure as F (fully loaded)). Remaining UPFs are either OFF or in power-saving mode. In conclusion, redundant hardware usage can be avoided, resulting in energy savings. Further AI/ML-based learning models can be incorporated in the core controller to understand traffic patterns, resource availability, and requirements. Such a learning-based framework for resource allocation in the core network is required to support diverse user and application trends and can also contribute to sustainability goals.  

\subsection{Energy-Aware ML-Driven Radio Access Network} 
\begin{figure}[h!]
\vspace{-0.4cm}
\centering
\includegraphics[width=0.8\columnwidth]{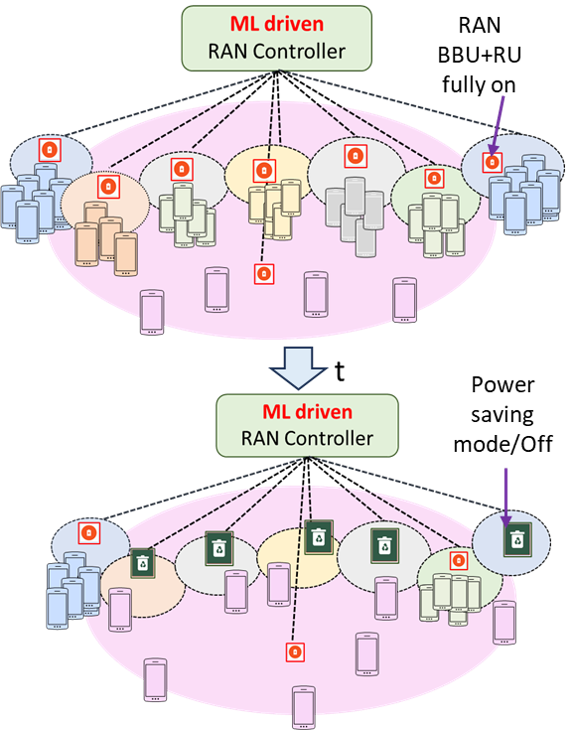}
\vspace{-0.4cm}
\caption{Energy-aware ML based dynamic resource allocation in RAN.}
\label{ran}
\vspace{-0.2cm}
\end{figure}

This use case highlights real-time resource allocation strategies for improving energy efficiency and sustainability in RAN  to enable support for massive connectivity requirements and diverse usage scenarios. In addition to the online-on-off-control,  where the RAN controller assigns active users to a few base stations that can effectively serve them, link-level optimizations and periodic sleep mode control at base stations can be implemented to reduce energy consumption in the access network. An illustration of the idea is shown in Figure \ref{ran}. Initially, all cells are ON to serve all the users in the area. After time $t$, the load conditions have changed, and only three cells will be ON to serve the active users. 
An additional strategy of dynamic selection of a suitable RAT, e.g., broadcasting RAT, may be particularly suitable for saving energy and radio resources when many users are accessing the same content. Further ML techniques can be implemented for resource allocation in ultra-dense mmWave/subTHz access networks. In such a deployment, we can dynamically and jointly choose user association, beam selection, and service schedule to minimize energy consumption while supporting the diverse quality of service (QoS) requirements. Such ML-based resource allocation schemes are a promising direction to achieve near-optimal performance with minimal resources.

\subsection{Service-Aware Adaptive Security Mechanism}
\begin{figure}[h!]
\vspace{-0.2cm}
\centering
\includegraphics[width=\columnwidth]{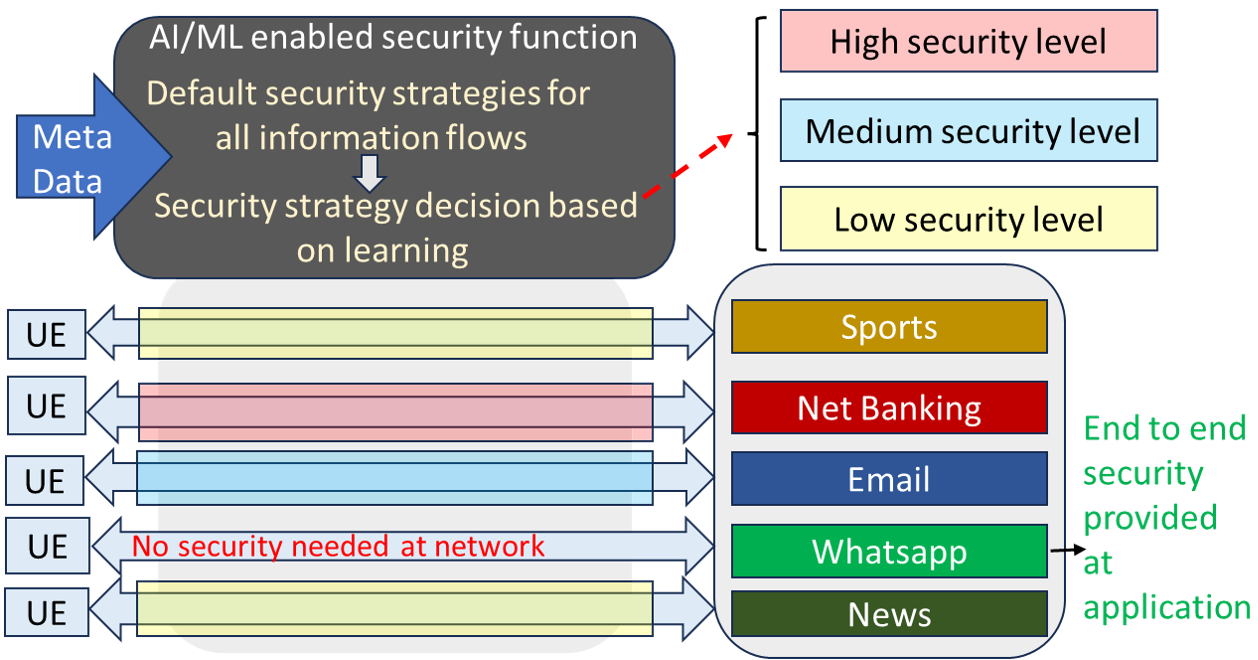}
\vspace{-0.4cm}
\caption{Service-aware adaptive security mechanism.}
\label{security}
\vspace{-0.2cm}
\end{figure}

This use case is for implementing a service-specific security strategy for individual flows. The underlying realization is that not all data flows necessitate the same level of security, allowing for varied security measures based on flow types, such as real-time streaming of a sporting event versus online financial transactions. Notably, current cellular networks lack differentiated security strategies, employing uniform measures for all flow types. Figure \ref{security} shows a broad-level idea of applying different service levels based on service type. For example, a messaging application (WhatsApp) applies end-to-end security at the application level. So there is no need to apply security into the network. In contrast, high security is needed to ensure banking-type services. Security strategy decisions can be AI/ML-enabled based on learnings from metadata and channel conditions. Learning techniques enable understanding data flow types and their security requirements, facilitating dynamically adjusted security strategies. Implementing adaptive security strategies can lead to energy and resource savings in network security implementation. 
\subsection{Energy-Aware   Multicasting}
\begin{figure}[h!]
\vspace{-0.2cm}
\centering
\includegraphics[width=\columnwidth]{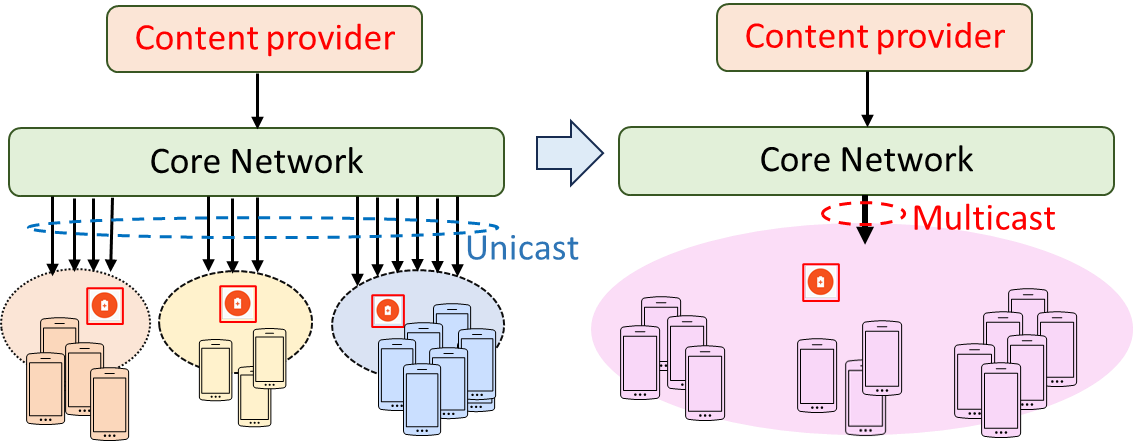}
\vspace{-0.2cm}
\caption{Switching to multicast delivery in place of unicast delivery when users are accessing same content.}
\label{multicast}
\vspace{-0.2cm}
\end{figure}

When multiple users are accessing the same content, multiple unicast data flows can be converted into a single multicast/broadcast data flow as shown in Figure~\ref{multicast}. For unicast data flows, energy consumption can linearly increase with the number of unicast sessions for UEs. In contrast, a multicast/broadcast session does not have such a relationship between energy consumption and the number of UEs \cite{wny}. A report concluded that the unicast method utilizes 24,000 kWh for streaming to 100,000 concurrent devices \cite{wny}. The broadcasting method only consumes 154.2 kWh for the same number of devices. Multicasting/broadcasting can help save energy in the RAN and core, as only a single data flow must be transported via UPF and RAN node to serve multiple users. Hence, an energy-aware   decision to use multicasting over unicasting can be sustainable for some specific types of services, such as live streaming. 
\section{Potential Benefits of Energy-aware   Networks}\label{benefits} There are several benefits of considering energy information-based approaches while designing the future networks, which can be classified from the network's and users' perspectives. The following subsections highlight some of those benefits.
\subsection{Network's Perspective}
\begin{itemize}
\item Overall energy saving: As sustainability is one of the prime goals for future networks (as per IMT 2030 requirements \cite{itu-M2160}), energy-saving oriented design principles like energy-aware   network decisions and energy optimization that consider joint goals, including energy consumption, can result in substantial energy savings. An essential benefit of deploying such approaches is energy saving at all levels in the network, and it is not a specific solution/approach for just a part of the network.
\item Priority to use renewable energy sources: Using the knowledge of energy source type as energy information for network decision can help prioritize renewable energy sources over conventional energy sources. Such prioritization can serve many benefits, like reduced carbon emissions and energy independence from grid energy. Further, it can reduce long-term operational costs and enhance operators' reputation towards environmental awareness.
\item Coverage extension to energy constraint areas: Introducing energy-aware   services can open a way for affordable and low-energy deployment solutions. With the aim of ``Connectivity everywhere and for everyone", there should be a focus on such deployment options with low energy services with low QoS, considering operators' benefits. These solutions can be used to provide coverage in the unconnected areas with energy constraints.
\item Efficiency improvement: Applying energy-aware   design principles with AI/ML technologies can result in smart energy management along with optimal network performance. 
\end{itemize}
\subsection{Users' Perspective}
\begin{itemize}
\item User satisfaction: Users can be environmentally sensitive and want to contribute to saving energy and increasing green energy usage. Providing energy information exposure and giving the option of choosing services based on energy is a choice for users to contribute to the environment. 
\item Cost benefits: The service's QoS, energy, service type, and cost are all dependent factors. Energy-based service selection can impact service cost for the user. If a user chooses a low-energy service, the user will have cost benefits, which is undoubtedly an advantage from the user's perspective.
\item Option for limiting maximum energy usage: For the users of the energy-constrained area, it is good to have a limit on energy usage at a granularity (per user, per service, etc.). It can help provide services in such areas, considering limited or unpredictable energy source availability (due to renewable energy sources). It is always better to have a ``service with low QoS" than ``no service," which can be made possible by limiting service-wise energy limits for users in areas with energy limitations. 
\end{itemize}

\section{Standardization Activities for Sustainability}\label{standards}
\subsection{3GPP}
There has been a lot of attention in 3GPP related to energy efficiency, energy metrics, and energy saving in mobile networks. Many Study Groups (SGs) have published specifications/reports and initiated work items in this direction, as listed in Table \ref{3gpp}. Starting from Release 15, a study report \cite{tr} covers system-wide energy-related Key Performance Indicators (KPIs) and architectural requirements to support
energy-saving capabilities at the network and equipment level. Key issues and respective solutions are studied in \cite{tr28813} related to the energy control framework and estimation methods for energy consumption in the network. Energy data collection requirements and procedures are detailed in \cite{ts28310}. Standardized energy-related KPIs associated with various network nodes in existing networks are available in \cite{ts28554}. Energy saving techniques in radio access networks are available in a study report from the 3GPP RAN group \cite{tr38864}.
\begin{table}[h!]
\caption{3GPP Specifications and Study Reports in different Domains}
\label{3gpp}
\centering
\begin{tabular}{|p{0.22\columnwidth} | p{0.68\columnwidth}|}
\hline
\textbf{Specification} & \textbf{Title} \\\hline \hline
\multicolumn{2}{|c|}{\textbf{Requirements}} \\ \hline 
TR 22.870 \cite{tr22870}& Study on 6G Use Cases and Service Requirements (Clause 5) \\ \hline
TS 22.261 \cite{ts22261} & Service requirements for the 5G system (Clause 6.15) \\ \hline
TR 22.882 \cite{tr22882} & Study on Energy Efficiency as a service criteria (Release 19)
\\\hline
TR 22.883 \cite{tr22883} & Study on Energy Efficiency as a service criteria Ph2 (Release 20)
\\\hline
\multicolumn{2}{|c|}{\textbf{System Architecture}} \\ \hline
TR 23.700-66 \cite{tr23700}  & Study on Energy Efficiency and Energy Saving
(Release 19) \\ \hline
TS 23.501 \cite{ts23501} & System architecture for the 5G
System (5GS); Stage 2 (Release 19)
\\\hline
\multicolumn{2}{|c|}{\textbf{Radio Access Network}} 
\\ \hline
RP-251395 \cite{ran-sid} & (Approved) Revised SID: Study on 6G Scenarios and requirements\\ \hline
TR 38.864 \cite{tr38864} & Study on network energy savings for
NR (Release 18) \\ \hline
\multicolumn{2}{|c|}{\textbf{Security}} \\ \hline
TR 33.766 \cite{tr33766}   & Study on security aspects of energy savings in 5G (Release 19)\\ \hline
\multicolumn{2}{|c|}{\textbf{Energy management, orchestration and parameters}} 
\\ \hline
TS 28.554 \cite{ts28554} & Management and orchestration;
5G end to end Key Performance Indicators (KPI) (Release 19)\\ \hline
TR 28.880 \cite{tr28880}  &  Study on energy efficiency and energy saving aspects of 5G networks and services (Release 19) \\ \hline
TS 28.310 \cite{ts28310} & Management and orchestration; Energy
efficiency of 5G (Release 18) \\ \hline
TR 28.813 \cite{tr28813} & Management and orchestration; Study
on new aspects of Energy Efficiency
(EE) for 5G (Release 17) \\ \hline
TR 21.866 \cite{tr} & Study on Energy Efficiency Aspects of
3GPP Standards (Release 15) \\ \hline
\end{tabular}
\end{table}
\par In last few 3GPP meetings, in SA1 and SA2 (requirements and architecture related 3GPP groups), ongoing Release 19 work items and newly initiated Release 20 study items, emphasize energy awareness in mobile networks by collecting energy-related information from network functions, exposing it securely to users and third parties (application service providers), and adjusting the Quality of Service (QoS) for data flows based on energy-related information \cite{tr22882, tr22883, tr23700}. Here, energy-related information includes per-slice and user (subscriber)  energy usage information for network functions and the ratio of the type of energy resources, conventional or renewable, used to serve a user.  It is also evident from the requirements of clause 6.15 of TS 22.261 \cite{ts22261} that significant enhancements are needed to collect and expose energy information within the network. The requirements related to modifying the service or adjusting the service quality based on energy-related information may require periodic collection and analysis. As a result of these discussions and approvals in the direction of energy information exposure and monitoring, a new function called Energy Information Function (EIF) is included as part of the 5G system architecture \cite{ts23501}, and related key issues and solutions for associated procedures are being discussed in \cite{tr23700}. All these developments are for the 5G network. From the 6G network's perspective, a new TR in SA1 \cite{tr22870} also has a clause titled ``System and Operational Aspects" that covers sustainability as part of its scope. In addition, a new SID is approved in the RAN group \cite{ran-sid}, which includes an objective of ``Energy efficiency and energy saving: both for network and device" for 6G. Considering the many open directions in 3GPP, there are many opportunities for contributions in different study groups.

\subsection{ITU}
ITU, a global standardization forum for telecommunication networks, plays a vital role in setting up guidelines for wireless and wireline networks. In the context of mobile networks, IMT-2030 requirements provided in an ITU recommendation \cite{itu-M2160} highlight sustainability in more than one clause, which is a significant motivational factor for 6G to consider sustainable design as a priority. Further, regarding specific work for energy-related aspects, a subgroup Q5 in SG5 ITU-T covers the scope related to ``Environmental efficiency of telecommunications/ICTs" \cite{ituq6}.
\begin{table}[h!]
\caption{ITU Recommendations (Published)}
\label{itu1}
\centering
\begin{tabular}{|p{0.3\columnwidth} | p{0.6\columnwidth}|}
\hline
\textbf{Recommendation} & \textbf{Title} \\ \hline
ITU-T L.1210 \cite{itu-L1210}& Sustainable power-feeding solutions for 5G networks \\ \hline
Series L Supplement 36 \cite{itu-S36} & Study on methods and metrics to evaluate energy efficiency for future 5G systems \\ \hline
ITU-T L.1331\cite{itu-L1331} & Assessment of mobile network energy efficiency 
\\\hline
ITU-T L.1350\cite{itu-L1350} & Energy efficiency metrics of a base station site
\\\hline
ITU-T L.1351\cite{itu-L1351} & Energy efficiency measurement methodology for base station sites
\\ \hline
Series L Supplement 43 \cite{itu-S43}& Smart energy saving of 5G base stations: Traffic forecasting and strategy optimization of 5G wireless network energy consumption based on artificial intelligence and other emerging technologies
\\ \hline
\end{tabular}
\end{table}
\begin{table}[h!]
\caption{Ongoing Work Items in ITU-T SG5 Q6 \cite{ituq6}}
\label{itu2}
\centering
\begin{tabular}{|p{0.2\columnwidth} | p{0.7\columnwidth}|}
\hline
\textbf{Work item} & \textbf{Title} \\ \hline
L.1210rev & Sustainable power-feeding solutions for 5G networks \\ \hline
L.1331rev & Assessment of mobile network energy efficiency 
\\\hline
L.1396 (ex L.MCI\_MIM) & Monitoring and Control Interface for Infrastructure Equipment (Power, Cooling and Building Environment Systems used in Telecommunication Networks) - ICT equipment power, energy and environmental parameters monitoring information model
\\\hline
L.1397 (ex L.MCI\_Bat) & Monitoring and control interface for infrastructure equipment (Power, Cooling and environment systems used in telecommunication networks) - Battery system with integrated control and monitoring information model)
\\ \hline
\end{tabular}
\end{table}
\par  A few published ITU recommendations (listed in the Table \ref{itu1}) are directly or indirectly related to energy aspects for mobile networks. Powering solutions and strategies for 5G core and RAN are proposed in \cite{itu-L1210}, including renewable energy solutions for 5G base stations and intelligent energy management in the mobile network. Energy efficiency metrics for mobile systems are detailed in \cite{itu-S36}. Further, assessment and measurement of energy efficiency-related metrics are covered in \cite{itu-L1331}. Whereas \cite{itu-L1350} and \cite{itu-L1351} are specific for the base station's energy metric and associated measurements. Considering the role of machine learning in smart energy saving, a supplement focuses on AI/ML-driven energy optimization support and methods for mobile networks \cite{itu-S43}.
\par Revision of a few of these recommendations is now open for contributions in ITU meetings as ongoing work items in ITU-T SG5 Q6, as listed in Table \ref{itu2}. In addition to contributions in open work items, many design directions are discussed in this survey paper that can be forwarded to ITU as new work item proposals.
\subsection{IEEE}
In the IEEE standardization forum, minimal work is available related to mobile network energy efficiency (as listed in Table \ref{ieee}). IEEE standard activities related to 5G and 6G are provided in \cite{ieee5g} and \cite{ieee6g}, respectively. 
\begin{table}[h!]
\caption{Few Relevant IEEE Standards}
\label{ieee}
\centering
\begin{tabular}{|p{0.3\columnwidth} | p{0.6\columnwidth}|}
\hline
\textbf{Standard/draft standard} & \textbf{Title} \\ \hline
IEEE 2061-2024\cite{ieee2061} & IEEE Standard for Architecture for Low Mobility Energy Efficient Network for Affordable Broadband Access
\\\hline
IEEE P1941 \cite{p1941} & Recommended Practice for Internet Grades of Service in Rural Areas \\  \hline
IEEE P1959 \cite{p1959} & Standard for Metrics for Energy Efficiency in Machine Learning Hardware Architectures and Systems
\\\hline
\end{tabular}
\end{table}

Related to 5G, there is an IEEE standard for Frugal 5G \cite{ieee2061}, that proposes an architecture for a low mobility energy efficient network for affordable broadband access. It has a limited focus on low-energy solutions for the middle-mile network. As per an article by IEEE Standard Association (SA), 6G Goals include sustainability aspects like increased usage of renewable energy, smart grids to optimize energy efficiency in the networks\cite{ieee6g-article}. However, IEEE SA contribution for 6G is centric to three areas \cite{ieee6g}: rural connectivity, consumer connectivity, and industry connectivity. In this direction, there is an ongoing draft standard on ``Recommended Practice for Internet Grades of Service in Rural Areas", which covers the classification of services based on mapping with QoS to meet the needs of rural areas  \cite{p1941}. To a certain extent, this proposal is aligned with the classification of energy-aware     services presented in the above section. Related to energy efficiency, there is another ongoing work in IEEE on ``Standard for Metrics for Energy Efficiency in Machine Learning Hardware Architectures and Systems" \cite{p1959}. This work is vital from the perspective of AI/ML integration's impact on the energy efficiency of the networks. In conclusion, IEEE is open to proposals for 6G networks, and there is a lot of scope to work on new draft standards in the area of energy aspects for mobile networks.

\section{Open Research Problems and Challenges} \label{open}
Based on ongoing discussions and extensive findings on energy efficiency in mobile networks, the following are some identified research directions that have limited discussions and now need attention from concerned communities to achieve future sustainability goals. 
\begin{itemize}
\item Explore feasible solutions for getting information and predicting renewable energy source consumption in real-time scenarios
\item Investigate procedures for energy monitoring and exposure to minimize procedural overhead
\item Analyze energy overhead due to energy data collection and analysis
\item AI/ML based solutions to overcome trade-off between overall energy saving and AI/ML energy overhead 
\item Methods to assure user consent for adjustment of QoS based on energy information
\end{itemize}
Above-mentioned points are the technical research directions to explore how optimal energy awareness support can be managed in the network. However, coordination and consent from all stakeholders are the biggest non-technical challenges. 
\section{Conclusion}\label{conclusion}
We present a multidimensional view of energy-related aspects for existing and future mobile networks. The need to enhance energy awareness in intelligent future mobile networks is discussed with the help of some highlights, such as prioritization of increased usage of renewable energy sources, green user satisfaction, connectivity gaps in energy-constrained areas, and AI/ML energy usage and energy-saving trade-off. Based on learning from ongoing activities in research communities and  standardization forums, design principles are proposed for energy-aware  6G mobile networks. The main directions of the new design focus on energy-aware network decisions, solutions for energy information exposure and monitoring in the network, and joint optimization of performance and energy using AI/ML. In addition, we present the latest standardization efforts in 3GPP, ITU, and IEEE on energy-affiliated approaches for 5G/6G mobile networks. We also provide insights into some open work items and proposals for future contributions. We conclude our survey by highlighting some open problems and challenges for designing an energy-aware     mobile network, especially the need for acceptable and feasible solutions to support a sustainable, secure, and modular energy-efficient network. Overall, we emphasize rethinking the design perspective for 6G networks by considering energy awareness into the design.
\section*{Acknowledgment}
We thank the Ministry of Electronics and Information Technology (MeitY), Government of India for supporting our work. Authors S. Moothedath and M. G. Bhat acknowledge the support from NSF-CNS 2415213.
\bibliographystyle{IEEEtran}
\bibliography{convergence.bib}
\end{document}